\documentclass[12pt]{article}
\usepackage{epsfig}
\usepackage{amsmath}
\usepackage{hhline}
\usepackage{amssymb}
\usepackage{times}
\usepackage{cite}
\newlength{\dinwidth}
\newlength{\dinmargin}
\begin{document}  
%%%%%%%%%%%%%%%% Pre-defined commands, you can use for the most obvious notations
\newcommand{\pom}{{I\!\!P}}
\newcommand{\reg}{{I\!\!R}}
\newcommand{\slowpi}{\pi_{\mathit{slow}}}
\newcommand{\fiidiii}{F_2^{D(3)}}
\newcommand{\fiidiiiarg}{\fiidiii\,(\beta,\,Q^2,\,x)}
\newcommand{\n}{1.19\pm 0.06 (stat.) \pm0.07 (syst.)}
\newcommand{\nz}{1.30\pm 0.08 (stat.)^{+0.08}_{-0.14} (syst.)}
\newcommand{\fiidiiiful}{F_2^{D(4)}\,(\beta,\,Q^2,\,x,\,t)}
\newcommand{\fiipom}{\tilde F_2^D}
\newcommand{\ALPHA}{1.10\pm0.03 (stat.) \pm0.04 (syst.)}
\newcommand{\ALPHAZ}{1.15\pm0.04 (stat.)^{+0.04}_{-0.07} (syst.)}
\newcommand{\fiipomarg}{\fiipom\,(\beta,\,Q^2)}
\newcommand{\pomflux}{f_{\pom / p}}
\newcommand{\nxpom}{1.19\pm 0.06 (stat.) \pm0.07 (syst.)}
\newcommand {\gapprox}
   {\raisebox{-0.7ex}{$\stackrel {\textstyle>}{\sim}$}}
\newcommand {\lapprox}
   {\raisebox{-0.7ex}{$\stackrel {\textstyle<}{\sim}$}}
\def\gsim{\,\lower.25ex\hbox{$\scriptstyle\sim$}\kern-1.30ex%
\raise 0.55ex\hbox{$\scriptstyle >$}\,}
\def\lsim{\,\lower.25ex\hbox{$\scriptstyle\sim$}\kern-1.30ex%
\raise 0.55ex\hbox{$\scriptstyle <$}\,}
\newcommand{\pomfluxarg}{f_{\pom / p}\,(x_\pom)}
\newcommand{\dsf}{\mbox{$F_2^{D(3)}$}}
\newcommand{\dsfva}{\mbox{$F_2^{D(3)}(\beta,Q^2,x_{I\!\!P})$}}
\newcommand{\dsfvb}{\mbox{$F_2^{D(3)}(\beta,Q^2,x)$}}
\newcommand{\dsfpom}{$F_2^{I\!\!P}$}
\newcommand{\gap}{\stackrel{>}{\sim}}
\newcommand{\lap}{\stackrel{<}{\sim}}
\newcommand{\fem}{$F_2^{em}$}
\newcommand{\tsnmp}{$\tilde{\sigma}_{NC}(e^{\mp})$}
\newcommand{\tsnm}{$\tilde{\sigma}_{NC}(e^-)$}
\newcommand{\tsnp}{$\tilde{\sigma}_{NC}(e^+)$}
\newcommand{\st}{$\star$}
\newcommand{\sst}{$\star \star$}
\newcommand{\ssst}{$\star \star \star$}
\newcommand{\sssst}{$\star \star \star \star$}
\newcommand{\tw}{\theta_W}
\newcommand{\sw}{\sin{\theta_W}}
\newcommand{\cw}{\cos{\theta_W}}
\newcommand{\sww}{\sin^2{\theta_W}}
\newcommand{\cww}{\cos^2{\theta_W}}
\newcommand{\trm}{m_{\perp}}
\newcommand{\trp}{p_{\perp}}
\newcommand{\trmm}{m_{\perp}^2}
\newcommand{\trpp}{p_{\perp}^2}
\newcommand{\alp}{\alpha_s}

\newcommand{\alps}{\alpha_s}
\newcommand{\sqrts}{$\sqrt{s}$}
\newcommand{\LO}{$O(\alpha_s^0)$}
\newcommand{\Oa}{$O(\alpha_s)$}
\newcommand{\Oaa}{$O(\alpha_s^2)$}
\newcommand{\PT}{p_{\perp}}
\newcommand{\JPSI}{J/\psi}
\newcommand{\sh}{\hat{s}}
\newcommand{\uh}{\hat{u}}
\newcommand{\MP}{m_{J/\psi}}
\newcommand{\PO}{I\!\!P}
\newcommand{\xbj}{x}
\newcommand{\xpom}{x_{\PO}}
\newcommand{\ttbs}{\char'134}
\newcommand{\xpomlo}{3\times10^{-4}}  
\newcommand{\xpomup}{0.05}  
\newcommand{\dgr}{^\circ}
\newcommand{\pbarnt}{\,\mbox{{\rm pb$^{-1}$}}}
\newcommand{\gev}{\,\mbox{GeV}}
\newcommand{\WBoson}{\mbox{$W$}}
\newcommand{\fbarn}{\,\mbox{{\rm fb}}}
\newcommand{\fbarnt}{\,\mbox{{\rm fb$^{-1}$}}}
\newcommand{\dsdx}[1]{$d\sigma\!/\!d #1\,$}
\newcommand{\eV}{\mbox{e\hspace{-0.08em}V}}
%
% Some useful tex commands
%
\newcommand{\qsq}{\ensuremath{Q^2} }
\newcommand{\gevsq}{\ensuremath{\mathrm{GeV}^2} }
\newcommand{\et}{\ensuremath{E_t^*} }
\newcommand{\rap}{\ensuremath{\eta^*} }
\newcommand{\gp}{\ensuremath{\gamma^*}p }
\newcommand{\dsiget}{\ensuremath{{\rm d}\sigma_{ep}/{\rm d}E_t^*} }
\newcommand{\dsigrap}{\ensuremath{{\rm d}\sigma_{ep}/{\rm d}\eta^*} }

%%% Dstar stuff
\newcommand{\dstar}{\ensuremath{D^*}}
\newcommand{\dstarp}{\ensuremath{D^{*+}}}
\newcommand{\dstarm}{\ensuremath{D^{*-}}}
\newcommand{\dstarpm}{\ensuremath{D^{*\pm}}}
\newcommand{\zDs}{\ensuremath{z(\dstar )}}
\newcommand{\Wgp}{\ensuremath{W_{\gamma p}}}
\newcommand{\ptds}{\ensuremath{p_t(\dstar )}}
\newcommand{\etads}{\ensuremath{\eta(\dstar )}}
\newcommand{\ptj}{\ensuremath{p_t(\mbox{jet})}}
\newcommand{\ptjn}[1]{\ensuremath{p_t(\mbox{jet$_{#1}$})}}
\newcommand{\etaj}{\ensuremath{\eta(\mbox{jet})}}
\newcommand{\detadsj}{\ensuremath{\eta(\dstar )\, \mbox{-}\, \etaj}}

% Journal macro
\def\Journal#1#2#3#4{{#1} {\bf #2} (#3) #4}
\def\NCA{\em Nuovo Cimento}
\def\NIM{\em Nucl. Instrum. Methods}
\def\NIMA{{\em Nucl. Instrum. Methods} {\bf A}}
\def\NPB{{\em Nucl. Phys.}   {\bf B}}
\def\PLB{{\em Phys. Lett.}   {\bf B}}
\def\PRL{\em Phys. Rev. Lett.}
\def\PRD{{\em Phys. Rev.}    {\bf D}}
\def\ZPC{{\em Z. Phys.}      {\bf C}}
\def\EJC{{\em Eur. Phys. J.} {\bf C}}
\def\CPC{\em Comp. Phys. Commun.}

%%%%%%%%%%%%% my new commands %%%%%%%%%%%%%%%%
\newcommand{\be}{\begin{equation}} 
\newcommand{\ee}{\end{equation}} 
\newcommand{\ba}{\begin{eqnarray}} 
\newcommand{\ea}{\end{eqnarray}}

\begin{titlepage}

\noindent
\begin{flushleft}
{\tt DESY 09-040    \hfill    ISSN 0418-9833} \\
{\tt June 2009}                  \\
\end{flushleft}

\vspace{2cm}
\begin{center}
\begin{Large}

{\bf Search for Excited Quarks in \begin{boldmath}$ep$\end{boldmath} Collisions at HERA \\}

\vspace{2cm}

H1 Collaboration

\end{Large}
\end{center}

\vspace{2cm}

\begin{abstract}
A search for excited quarks is performed using the full  $e^{\pm}p$~data sample collected by the H1 experiment at HERA, corresponding to a total integrated luminosity of $475$~pb$^{-1}$.
The electroweak decays of excited quarks ${q}^{*}\rightarrow{q}{\gamma}$, ${q}^{*}\rightarrow{q}Z$ and ${q}^{*}{\rightarrow}q W$ with subsequent hadronic or leptonic decays of the $W$ and $Z$ bosons are considered.
No evidence for first generation excited quark production is found. 
Mass dependent exclusion limits on $q^*$ production cross sections and on the ratio $f/{\Lambda}$ of the coupling to the compositeness scale are derived within gauge mediated models.
These limits extend the excluded region compared to previous excited quark searches.
\end{abstract}

\vspace{1.5cm}

\begin{center}
Accepted by \PLB
\end{center}

\end{titlepage}

%          THE PAPER DRAFTS HAVE NO AUTHORLIST
%
%          FOR PAPER ISSUED FOR THE FINAL READING 
%          COPY THE AUTHOR AND INSTITUTE LISTS 
%          INTO YOUR AREA
%
% from /h1/iww/ipublications/h1auts.tex 
%          AND UNCOMMENT THE NEXT THREE LINES 
%
\begin{flushleft}

%-- H1AUTS Author list by names 
%-- Status: Mon Feb  9 09:07:51 CET 2009  Number of authors = 253 

F.D.~Aaron$^{5,49}$,           %BUCH-PD        11/06           Aaron               
C.~Alexa$^{5}$,                %BUCH-PD        06/06           Alexa               
K.~Alimujiang$^{11}$,          %DESY-PD        07/08           Alimujiang          
V.~Andreev$^{25}$,             %LPI -PD        8/88            Andreev             
B.~Antunovic$^{11}$,           %DESY-LEFT      12/08           Antunovic           
A.~Asmone$^{33}$,              %ROME-ST        07/2            Asmone              
S.~Backovic$^{30}$,            %PODG-PD        03/2            Backovic            
A.~Baghdasaryan$^{38}$,        %YERE-PD        09/03           Baghdasaryana       
E.~Barrelet$^{29}$,            %PARI-PD        11/99           Barrelet            
W.~Bartel$^{11}$
% ...

%-- H1 Institutes 
\bigskip{\it
 $ ^{1}$ I. Physikalisches Institut der RWTH, Aachen, Germany$^{ a}$ \\
% ...
$ ^{5}$ National Institute for Physics and Nuclear Engineering (NIPNE) ,
          Bucharest, Romania \\
% ...
$ ^{11}$ DESY, Hamburg, Germany \\
% ...
$ ^{25}$ Lebedev Physical Institute, Moscow, Russia$^{ e}$ \\
% ...
$ ^{29}$ LPNHE, Universit\'{e}s Paris VI and VII, IN2P3-CNRS,
          Paris, France \\
 $ ^{30}$ Faculty of Science, University of Montenegro,
          Podgorica, Montenegro$^{ e}$ \\
% ...
$ ^{33}$ Dipartimento di Fisica Universit\`a di Roma Tre
          and INFN Roma~3, Roma, Italy \\
% ...
$ ^{38}$ Yerevan Physics Institute, Yerevan, Armenia \\
% ...
$ ^{49}$ Also at Faculty of Physics, University of Bucharest,
          Bucharest, Romania \\
% ...
$ ^a$ Supported by the Bundesministerium f\"ur Bildung und Forschung, FRG,
      under contract numbers 05 H1 1GUA /1, 05 H1 1PAA /1, 05 H1 1PAB /9,
      05 H1 1PEA /6, 05 H1 1VHA /7 and 05 H1 1VHB /5 \\
% ...
$ ^e$ Supported by the Deutsche Forschungsgemeinschaft \\
% ...
}

\end{flushleft}
%
% Please not that the author list may need re-formatting.

\newpage

%%%%%%%%%%%%%%%%%%%%%%%%%%%%%%%%%%%%%%%%%%%%%%%%%%%%%%%%%%%%
\section{Introduction}
%%%%%%%%%%%%%%%%%%%%%%%%%%%%%%%%%%%%%%%%%%%%%%%%%%%%%%%%%%%%%

The three-family structure and mass hierarchy of the known fermions is one of the most puzzling characteristics of the Standard Model (SM) of particle physics.
Attractive explanations are provided by models assuming a composite structure of quarks and leptons~\cite{Harari:1982xy}.
The existence of excited states of leptons and quarks is a natural consequence of these models and their discovery would be a convincing evidence of a new scale of matter.
Electron\footnote{In this paper the term ``electron'' is used generically to refer to both electrons and positrons, unless otherwise stated.}-proton interactions at high energies provide the opportunity to search for excited states of first generation fermions. 
For instance, excited quarks ($q^*$) could be singly produced through the exchange of a $\gamma$ or a $Z$ boson. 

In this paper a search for excited quarks using the complete $e^{\pm}p$  HERA collider data of the H1 experiment is presented.
Electroweak decays of the excited quark into a SM quark and a gauge boson ($\gamma$, $W$ and $Z$) are considered and both hadronic and leptonic decays of the $W$ and $Z$ bosons are analysed.

The data were recorded at an electron beam energy of $27.6$~GeV and proton beam energies of $820$~GeV or $920$~GeV, corresponding to centre-of-mass energies $\sqrt{s}$ of $301$~GeV and $319$~GeV, respectively.
The total integrated luminosity of the data sample is $475$~pb$^{-1}$. 
The data comprise $184$~pb$^{-1}$ recorded in $e^-p$ collisions and $291$~pb$^{-1}$ in $e^+p$ collisions, of which $35$~pb$^{-1}$ were recorded at \mbox{$\sqrt{s} = 301$~GeV}. 
With more than a twelve-fold increase in statistics and a higher centre-of-mass energy, this analysis supersedes the result of previous searches for excited quarks at HERA by the H1~\cite{Adloff:2000gv} and ZEUS~\cite{Chekanov:2001xk} Collaborations and is complementary to exclusion limits obtained at the LEP collider~\cite{Abreu:1998jw} and at the Tevatron~\cite{Abe:1993sv,Abe:1997hm,Abazov:2003tj,Aaltonen:2008dn}.
The analysis also complements searches for first generation excited neutrinos~\cite{Aaron:2008xe} and electrons~\cite{Collaboration:2008cy} at HERA.

%%%%%%%%%%%%%%%%%%%%%%%%%%%%%%%%%%%%%%%%%%%%%%%%%%%%%%%%%%%%%
\section{Phenomenology}
%%%%%%%%%%%%%%%%%%%%%%%%%%%%%%%%%%%%%%%%%%%%%%%%%%%%%%%%%%%%%

In the present study a model~\cite{Hagiwara:1985wt,Baur:1989kv,Boudjema:1992em} is considered in which excited fermions are assumed to have spin  $1/2$ and weak isospin $1/2$.
The left-handed and right-handed components of the excited fermions form weak iso-doublets $F_L^*$ and $F_R^*$.
Interactions between excited and ordinary fermions may be mediated by gauge bosons, as described by the effective Lagrangian~\cite{Baur:1989kv,Boudjema:1992em}:

\be
{\cal L}_{int.} = \frac{1}{2\Lambda}{\bar{F^{*}_{R}}} \; {{\sigma}^{\mu\nu}} \left[ gf\frac{\tau^a}{2}{W_{\mu\nu}^a}+g'f'\frac{Y}{2}B_{\mu\nu} + g_s f_s \frac{\lambda^a}{2} G^a_{\mu\nu} \right] \; {F_{L}} + h.c. \; .
\label{eq:lagrangian}
\ee

Only the right-handed component of the excited fermions $F_R^*$ is allowed to couple to light fermions $F_L$, in order to protect the light leptons from radiatively acquiring a large anomalous magnetic moment~\cite{Brodsky:1980zm,Renard:1982ij}.
The matrix ${{\sigma}^{\mu\nu}}$ is the covariant bilinear tensor, $W_{{\mu}{\nu}}^a$,  $B_{{\mu}{\nu}}$ and $G^a_{\mu\nu}$ are the field-strength tensors of the SU($2$), U($1$) and SU($3$)$_{C}$ gauge fields, $\tau^a$, $Y$ and $\lambda^a$ are the Pauli matrices, the weak hypercharge operator and the Gell-Mann matrices.
The standard electroweak and strong gauge couplings are denoted by $g$, $g'$ and $g_s$, respectively. 
The parameter $\Lambda$ has units of energy and can be regarded as the compositeness scale which reflects the range of the new confinement force. 
The constants $f$, $f'$ and $f_s$ are coupling parameters associated with the three gauge groups and are determined by the yet unknown composite dynamics. 

\begin{figure}[htbp] 
  \begin{center}
\includegraphics[height=6cm]{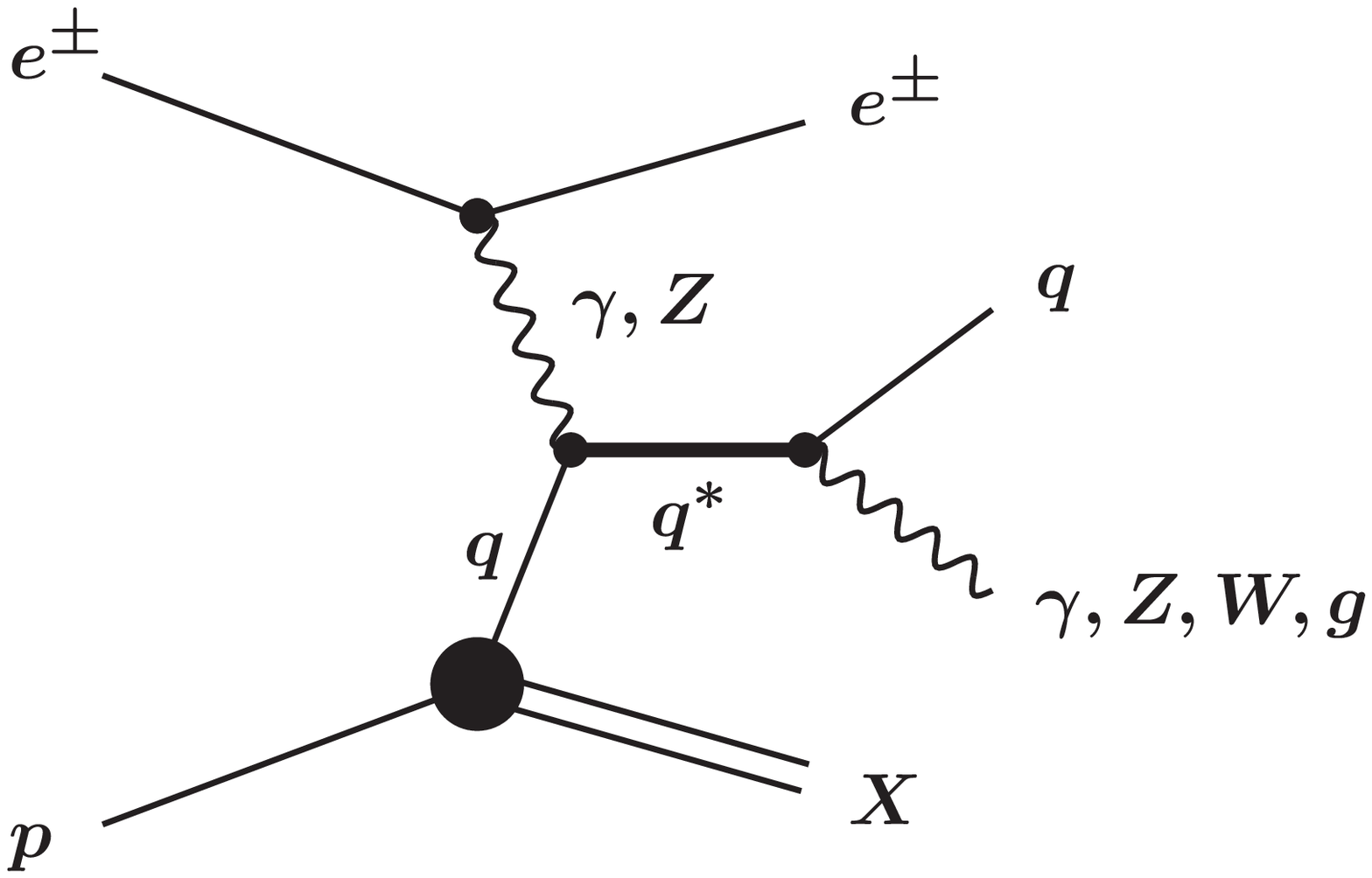}\put(-52,-2) {{\bf (a)}}
\includegraphics[height=6cm]{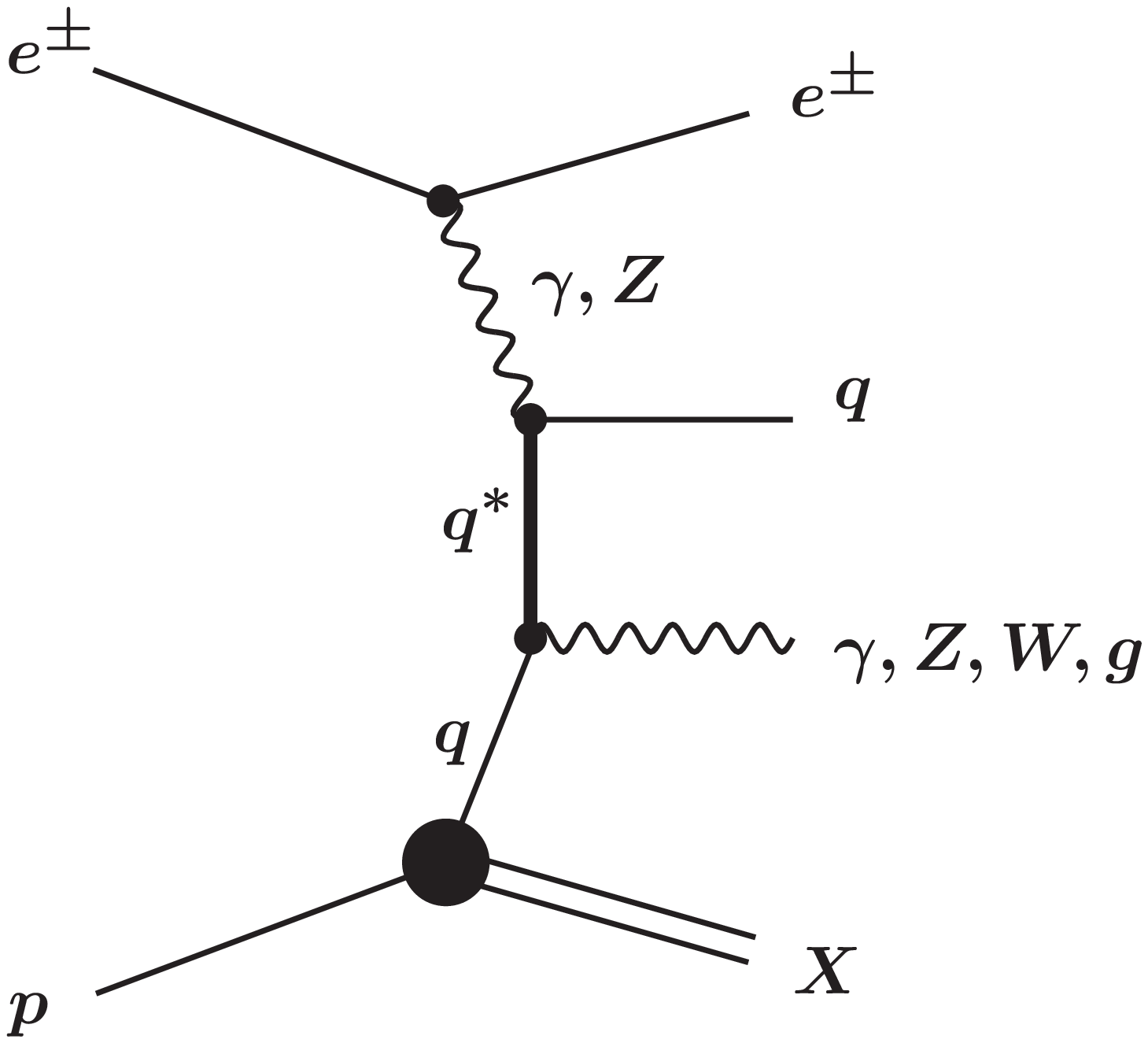}\put(-44,-2) {{\bf (b)}}\\
  \end{center}
 \caption{Diagrams for the production and decay of excited quarks in $ep$ collisions.}
 \label{fig:Diag}  
 \end{figure}

Following this model of gauge mediated interactions, excited quarks may be resonantly produced in $ep$ collisions through a gauge boson exchange between the incoming electron and a quark of the proton (see figure~\ref{fig:Diag}(a)).
%
%Differently to excited leptons~\cite{Aaron:2008xe,Collaboration:2008cy}, the cross section for the exchange of excited quarks in the $u$-channel (figure~\ref{fig:Diag}(b)) is non-negligible for the high $q^*$ masses and low values of $\Lambda$ probed in this analysis.
%
The exchange of excited quarks in the $u$-channel (figure~\ref{fig:Diag}(b)) is also 
possible for the high $q^*$ masses and low values of $\Lambda$ probed in this analysis.
For example, for $\Lambda = 50$~GeV, $f = f' = 1$ and an excited quark of mass $M_{q^*} = 300$~GeV, the $u$-channel production cross section is $0.016$~pb while the resonant production cross section is $0.27$~pb.

The excited quark may decay into a quark and a gauge boson via $q^* {\rightarrow} q\gamma$, $q^* {\rightarrow} q W$,  $q^* {\rightarrow} q Z$ and $q^* {\rightarrow} q g$.
For a given $q^*$ mass value and assuming a numerical relation between $f$, $f'$ and $f_s$, the $q^*$ branching ratios are fixed and the production cross section depends only on $f/\Lambda$.
Only $\gamma$, $W$ and $Z$ decays of the $q^*$ are considered in the present study.
In this analysis, the assumptions are made that the coupling parameters  $f$ and $f'$ are of comparable strength, with the relationship $f = f'$, and that $f_s = 0$.
These assumptions lead to results which are complementary to the $q^*$ searches performed at the Tevatron~\cite{Abe:1993sv,Abe:1997hm,Abazov:2003tj,Aaltonen:2008dn}, since at a $p\bar{p}$ collider excited quarks are dominantly produced in a quark-gluon fusion mechanism, which requires $f_s \neq 0$.
The effect of non-zero values of $f_s$ is also studied in the present analysis.

%%%%%%%%%%%%%%%%%%%%%%%%%%%%%%%%%%%%%%%%%%%%%%%%%%%%%%%%%%%%%
\section{Simulation of Signal and Background Processes}
%%%%%%%%%%%%%%%%%%%%%%%%%%%%%%%%%%%%%%%%%%%%%%%%%%%%%%%%%%%%%

A Monte Carlo (MC) program developed for this analysis is used for the calculation of the $q^*$ production cross section and the simulation of signal events.
The events are simulated using the cross section calculated from the Lagrangian described in equation~(\ref{eq:lagrangian}) using the \mbox{CompHEP} program~\cite{comphep}.
Both resonant $q^*$ production and $u$-channel exchange processes, as well as their interference are included.
Initial state radiation of a photon from the incident electron is included using the Weizs\"acker-Williams approximation~\cite{Berger:1986ii}. 
The proton parton densities are taken from the CTEQ5L~\cite{Pumplin:2002vw} parametrisation and are evaluated at the scale $\sqrt{\hat{s}}=\sqrt{sx}$, where $x$ is the momentum fraction of the proton carried by the interacting quark.
The parton shower approach~\cite{Sjostrand:2000wi} is applied in order to simulate Quantum Chromodynamics (QCD) corrections in the initial and final states. Hadronisation is performed using Lund string fragmentation as implemented in PYTHIA~\cite{Sjostrand:2000wi}.
In the MC generator the full transition matrix including both $q^*$ production and decay is implemented. 
This is important if the natural width of the $q^*$ is large, which is typically the case at high mass. 
In order to incorporate the width effects in the signal efficiency determination, events are generated with the coupling $f/\Lambda$ corresponding, for each $q^*$ mass, to the expected boundary of the probed domain in the plane defined by $M_{q^*}$ and  $f/\Lambda$.
Excited quarks will be searched for in the $q\gamma$, $q{q}{\bar{q}}$, $q e \nu$, $q \mu \nu$, $q e e$ and $q \mu \mu$ final states.
The SM background processes that may mimic the $q^*$ signal are prompt photon production, neutral current (NC) deep inelastic scattering (DIS), photoproduction, single $W$ boson production and lepton pair production. 

The RAPGAP~\cite{Jung:1993gf} event generator, which implements the Born, QCD Compton and Boson Gluon Fusion matrix elements, is used to model NC DIS events. 
The QED radiative effects arising from real photon emission from both the incoming and outgoing electrons are simulated using the HERACLES~\cite{Kwiatkowski:1990es} program. 
Direct and resolved photoproduction of jets and prompt photon production are simulated using the PYTHIA event generator. 
The simulation is based on Born level scattering matrix elements with radiative QED corrections. 
In RAPGAP and PYTHIA, jet production from higher order QCD radiation is simulated using leading logarithmic parton showers and hadronisation is modelled with Lund string fragmentation.
The leading order MC prediction of NC DIS and photoproduction processes with two or more high transverse momentum jets is scaled by a factor of $1.2$ to account for the incomplete description of higher orders in the MC generators~\cite{Adloff:2002au,Aktas:2004pz}. 
Charged current DIS events are simulated using the DJANGO~\cite{Schuler:yg} program, which includes first order leptonic QED radiative corrections based on HERACLES. The production of two or more jets in DJANGO is accounted for using the colour dipole model~\cite{Lonnblad:1992tz}. 
Contributions from elastic and quasi-elastic QED Compton scattering are simulated with the WABGEN~\cite{Berger:kp} generator. 
Contributions arising from the production of single $W$ bosons and multi-lepton events are modelled using the EPVEC~\cite{Baur:1991pp} and GRAPE~\cite{Abe:2000cv} event generators, respectively.

Generated events are passed through a GEANT~\cite{Brun:1987ma} based simulation of the H1 apparatus, which takes into account the actual running conditions of the data taking, and are reconstructed and analysed using the same program chain as is used for the data.

%%%%%%%%%%%%%%%%%%%%%%%%%%%%%%%%%%%%%%%%%%%%%%%%%%%%%%%%%%%%%
\section{Experimental Conditions}
%%%%%%%%%%%%%%%%%%%%%%%%%%%%%%%%%%%%%%%%%%%%%%%%%%%%%%%%%%%%%

A detailed description of the H1 detector can be found in \cite{Abt:h1_1}.
Only the detector components relevant to the
present analysis are briefly described here.  
The origin of the H1 coordinate system is the nominal $ep$ interaction point, with the direction of the proton beam defining the positive $z$-axis (forward region). Transverse momentum ($P_T$) is measured in the $x$--$y$ plane. The pseudorapidity $\eta$ is related to the polar angle $\theta$ by $\eta = -\ln \, \tan (\theta/2)$.
The Liquid Argon (LAr) calorimeter~\cite{Andrieu:1993kh} is used to measure energy and direction of electrons, photons and hadrons. It covers the polar angle range
$4^\circ < \theta < 154^\circ$ with full azimuthal acceptance.
Electromagnetic shower energies are measured with a precision of
$\sigma (E)/E = 12\%/ \sqrt{E/\mbox{GeV}} \oplus 1\%$ and hadronic energies
with $\sigma (E)/E = 50\%/\sqrt{E/\mbox{GeV}} \oplus 2\%$, as measured in test beams~\cite{Andrieu:1994yn,Andrieu:1993tz}.
In the backward region, energy measurements are provided by a lead/scintillating-fibre (SpaCal) calorimeter\footnote{This device was installed in 1995, replacing a lead-scintillator ``sandwich'' calorimeter~\cite{Abt:h1_1}.}~\cite{Appuhn:1996na} covering the angular range $155^\circ < \theta < 178^\circ$.
The central ($20^\circ < \theta < 160^\circ$) and forward ($7^\circ < \theta < 25^\circ$)  tracking detectors are used to
measure charged particle trajectories, to reconstruct the interaction
vertex and to complement the measurement of hadronic energy.
The innermost proportional chamber CIP  ($9^\circ < \theta < 171^\circ$) is used to veto charged particles for the identification of photons.
The LAr calorimeter and inner tracking detectors are enclosed in a superconducting magnetic
coil with a field strength of $1.16$~T.
From the curvature of charged particle trajectories in the magnetic field, the central tracking system provides transverse momentum measurements with a resolution of $\sigma_{P_T}/P_T = 0.5\% \; P_T / \rm{GeV} \oplus 1.5\%$~\cite{Kleinwort:2006zz}.
The return yoke of the coil is the outermost part of the central detector and is
equipped with streamer tubes forming the central muon detector
($4^\circ < \theta < 171^\circ$).
In the forward region of the detector ($3^\circ < \theta < 17^\circ$) a set of
drift chambers detects muons and measures their momenta using an iron toroidal magnet.
The luminosity is determined from the rate of the Bethe-Heitler process $ep {\rightarrow} ep \gamma$,
measured using a photon detector located close to the beam pipe at $z=-103~{\rm m}$, in the backward direction.

%%%%%%%%%%%%%%%%%%%%%%%%%%%%%%%%%%%%%%%%%%%%%%%%%%%%%%%%%%%%%
\section{Data Analysis}\label{sec:anal}
%%%%%%%%%%%%%%%%%%%%%%%%%%%%%%%%%%%%%%%%%%%%%%%%%%%%%%%%%%%%%

The triggers employed for collecting the data used in this analysis are based on the detection of electromagnetic and hadronic energy deposits or missing transverse energy in the LAr calorimeter~\cite{Adloff:2003uh}.
For events with missing transverse energy of $20$~GeV, the trigger efficiency is $\sim$~$90$\% and increases to above $95$\% for missing transverse energy above $30$~GeV.
Events containing an electromagnetic deposit (electron or photon) with an energy greater than $10$~GeV are triggered with an efficiency close to $100$\%~\cite{nikiforov}.
Events with two or three jets of transverse momentum larger than $20$~GeV are triggered with an efficiency of nearly $100$\%.

In order to remove background events induced by cosmic showers and other non-$ep$ sources, the event vertex is required to be reconstructed within $35$~cm in $z$ of the nominal interaction point. In addition, topological filters and timing vetoes are applied~\cite{negri}.

In a first analysis step, calorimetric energy deposits and tracks of the event are used to look for electron, photon and muon candidates.
Electron and photon candidates are characterised by compact and isolated electromagnetic showers in the LAr calorimeter.  
The identification of muon candidates is based on a track measured in the inner tracking systems associated with signals in the muon detectors~\cite{Andreev:2003pm,Aaron:2008jh}.
Calorimeter energy deposits and tracks not previously identified as electron, photon or muon candidates are used to form combined cluster-track objects, from which the hadronic energy is reconstructed~\cite{matti,benji}.
Jet candidates are reconstructed, with a minimum transverse momentum of $2.5$~GeV, from these combined cluster-track objects using an inclusive $k_T$ algorithm~\cite{Ellis:1993tq,Catani:1993hr} with a $P_T$ weighted recombination scheme in which the jets are treated as massless.
The missing transverse momentum  $P_T^{\rm{miss}}$ of the event is derived from all detected particles and energy deposits in the event.
In events with large $P_T^{\rm{miss}}$, a neutrino candidate is reconstructed.  
The four-vector of this neutrino candidate is calculated assuming  transverse momentum conservation and the relation $\sum_i (E^i - P_{z}^{i}) + (E^\nu - P_{z}^{\nu}) = 2 E^0_e = 55.2$~GeV, where the sum runs over all detected particles; $P_{z}$ is the momentum along the proton beam axis and $E^0_e$ is the electron beam energy. The later relation assumes that no significant losses are present in the electron beam direction.

In a second step, additional requirements are applied to ensure a clear  identification of particles.
For electrons and photons, the hadronic energy within a distance in the pseudorapidity-azimuth $(\eta - \phi)$ plane $R=\sqrt{\Delta \eta^2 + \Delta \phi^2} < 0.5$ around the electron (photon) is required to be below $3$\% of the electron (photon) energy.
Furthermore, each electron (photon) must be isolated from jets by a minimum distance in pseudorapidity-azimuth of $R > 0.5$ to any jet axis.
In the polar angle region $\theta^e >35^\circ$ electrons must be associated to a charged track and be isolated from any other track by a minimum distance of $R> 0.5$.
In the central region (${\theta}^{\gamma} > 20^{\circ}$), photons are selected only if no track points to the electromagnetic cluster neither within a distance of closest approach (DCA) of $30$~cm nor within $R <0.5$.
In the forward region (${\theta}^{\gamma} < 20^{\circ}$) only photon candidates with no good quality track with a DCA to the cluster below $12$~cm are accepted. In this region, the calorimetric isolation of the photon candidate is tightened by requiring that the hadronic energy within $R < 1$, instead of $R<0.5$, around the photon be below $3$\% of the photon energy.
In addition, it is required that no hit in the CIP be associated to the photon candidate.
A muon should have no more than $5$~GeV deposited in a
cylinder, centred on the muon track direction, of radius $25$~cm and $50$~cm in the electromagnetic and hadronic sections of the LAr calorimeter, respectively.
Additionally, the muon is required to be separated from the closest jet and from any track by $R > 1$ and  $R > 0.5$, respectively.
Specific selection criteria applied in each decay channel are presented in the following subsections.

\subsection{\boldmath $q\gamma$ Resonance Search}

The signature of the $q^* {\rightarrow} q \gamma$ decay channel consists of one high $P_T$ isolated electromagnetic cluster and one high $P_T$ jet. 
SM background arises from radiative NC DIS and prompt photon events.
As decay products of a massive particle have large transverse momenta and are boosted in the forward region, events are selected with a photon with transverse momentum \mbox{$P_T^\gamma >$ $35$~GeV} in a polar angle range \mbox{$5^{\circ} < {\theta}^{\gamma} < 90^{\circ}$}. The events are required to have at least one jet in the polar angle range  $5^{\circ} < {\theta}^{\rm{jet}} < 80^{\circ}$ with a transverse momentum larger than $20$~GeV.
Photoproduction background events typically have low  values of the Bjorken scaling variable, $x_h$, calculated from the hadronic system using the Jacquet-Blondel method~\cite{Adloff:1999ah,JBmethod}.
Their contribution is reduced by a factor of two by requiring $x_h > 0.1$.
Finally, to further reduce the background from NC DIS, it is required that no electromagnetic deposit with an energy larger than $10$~GeV with an associated track is present in the LAr.

After this selection, $44$  events are found in the data compared to a SM expectation of $46$~$\pm$~$8$ events. 
The errors on the SM prediction include model and experimental systematic errors added in quadrature (see section \ref{sec:sys_err}).
The remaining dominant SM background contributions are prompt photon ($66$\%) and radiative NC DIS ($26$\%) events.
The invariant mass of the $q^*$ candidate is calculated from the four-vectors of the photon and the jet candidate of highest $P_T$.
The invariant mass distribution of the $q^*$ candidates and the SM background expectation is presented in figure~\ref{fig:Mass}(a).
From Monte Carlo studies, the selection efficiency is $35$\% for $M_{q^*} = 120$~GeV, increasing to $45$\% for \mbox{$M_{q^*} = 260$~GeV}.
The total width of the reconstructed $q^*$ mass distribution is $6$~GeV for a generated $q^*$ mass of $120$~GeV, increasing to $12$~GeV for a $q^*$ mass of $260$~GeV.

\subsection{\boldmath $q{q}{\bar{q}}$  Resonance Search}

The signatures of the ${q}^{*} {\rightarrow} q W {\rightarrow} q q \bar{q}$ and  ${q}^{*} {\rightarrow} q Z {\rightarrow} q q \bar{q}$ decay channels are similar to each other and consist of three high transverse momentum jets.
The SM background is dominated by multi-jet photoproduction and NC DIS events.
Events are selected with at least three jets in the polar angle range $5^{\circ} < \theta^{\rm{jet}} < 120^{\circ}$ with transverse momenta larger than $50$, $30$ and $15$~GeV, respectively.
In each event, a $W$ or $Z$ boson candidate is reconstructed from the combination of the two jets with an invariant mass closest to the nominal $W$ or $Z$ boson mass. 
The reconstructed mass of the $W$ or $Z$ candidate is required to be larger than $70$~GeV and smaller than $100$~GeV.
From MC studies, in decays of $q^*$ of large mass, the highest $P_T$ jet often does not originate from the boson decay. Therefore, only events in which the highest $P_T$ jet is not associated to the $W$ or $Z$ boson candidate are selected. 
This requirement is particularly effective in suppressing the photoproduction background at high $q^*$ masses. However, it reduces the $q^*$ selection efficiency at low masses.

After this selection, $341$ events are observed while $326$ $\pm$ $78$ are expected from the SM.
The remaining dominant SM background contributions are photoproduction ($52$\%) and NC DIS ($39$\%) events.
The invariant mass of the $q^*$ candidate is calculated from the highest $P_T$ jet and $W$ or $Z$ candidate four-vectors.
The invariant mass distributions of the $q^*$ candidates and of the SM background are presented in figure~\ref{fig:Mass}(b).
The selection efficiency in this channel is  $5$\% for $M_{q^*} = 120$~GeV, increasing to $35$\% for $M_{q^*} = 160$~GeV and to $55$\% for $M_{q^*} = 260$~GeV.
The total width of the reconstructed $q^*$ mass distribution is $11$~GeV for a generated $q^*$ mass of $120$~GeV, increasing to $25$~GeV for a $q^*$ mass of $260$~GeV.

\subsection{\boldmath $q e \nu$ and $q \mu \nu$ Resonance Searches}

The signature of the ${q}^{*} {\rightarrow} {q} W {\rightarrow} qe\nu$ and ${q}^{*} {\rightarrow} {q} W {\rightarrow} q\mu\nu$ decay channels consists of one energetic lepton, a prominent jet and missing transverse momentum.
Events with this topology correspond in the SM to single $W$ production~\cite{Aaron:2009wp}.
Other SM background processes that may mimic the signature through misidentification or mismeasurement are NC and CC DIS, photoproduction and lepton pair production.

\begin{figure}[htbp] 
%   \begin{center}
\includegraphics[width=.5\textwidth]{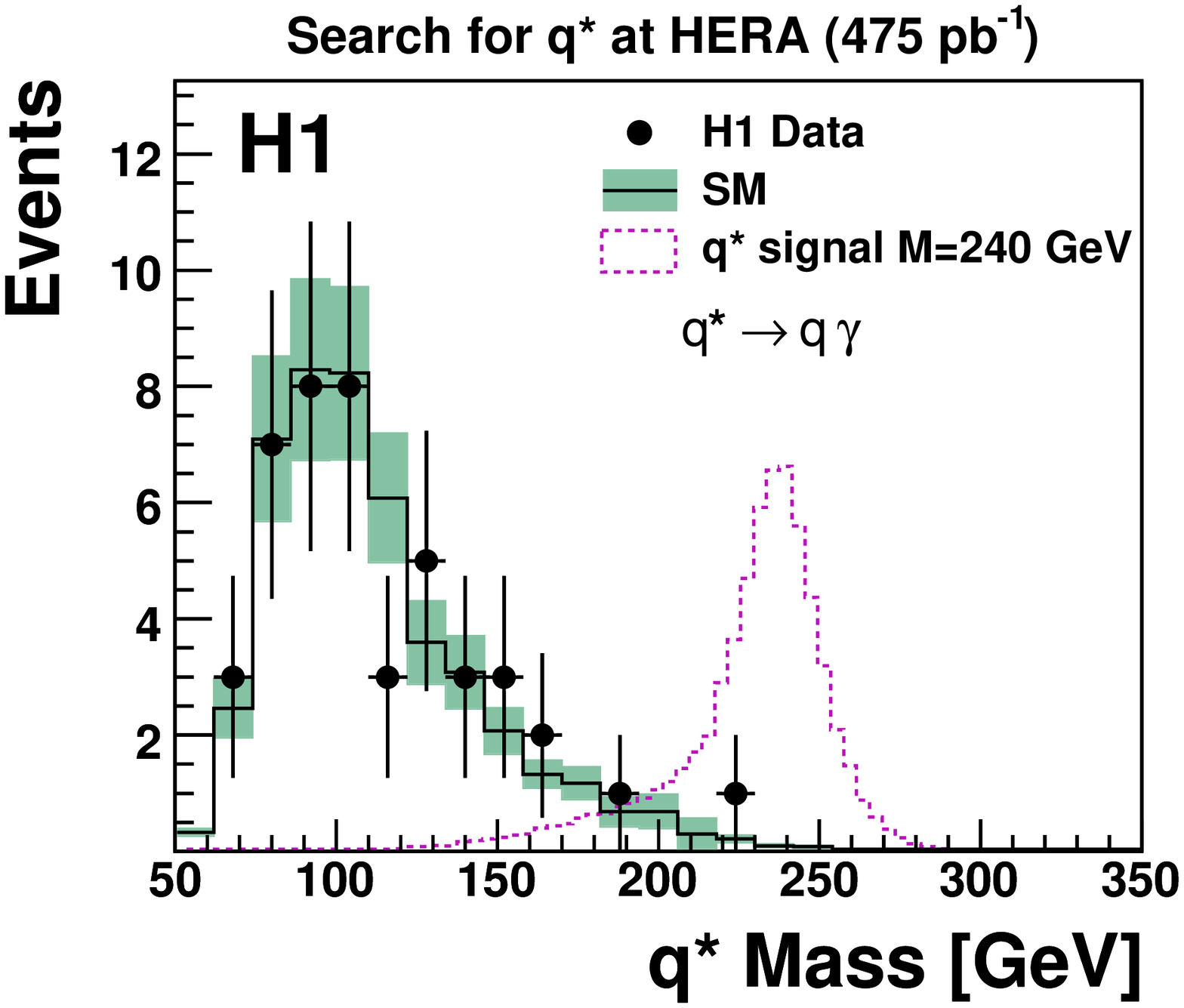}\put(-12,37) {{\bf (a)}}
\includegraphics[width=.5\textwidth]{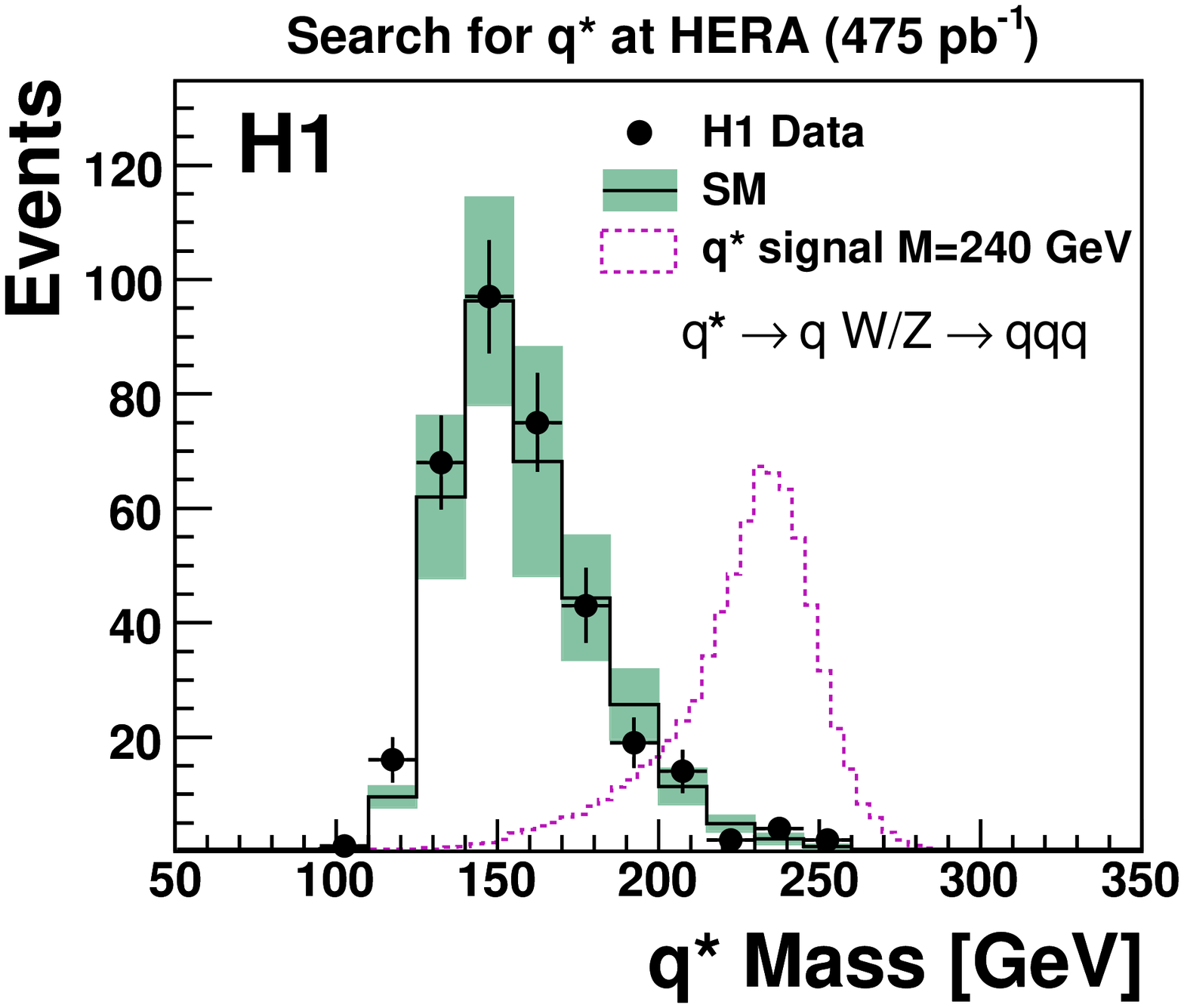}\put(-12,37) {{\bf (b)}}\\
\includegraphics[width=.5\textwidth]{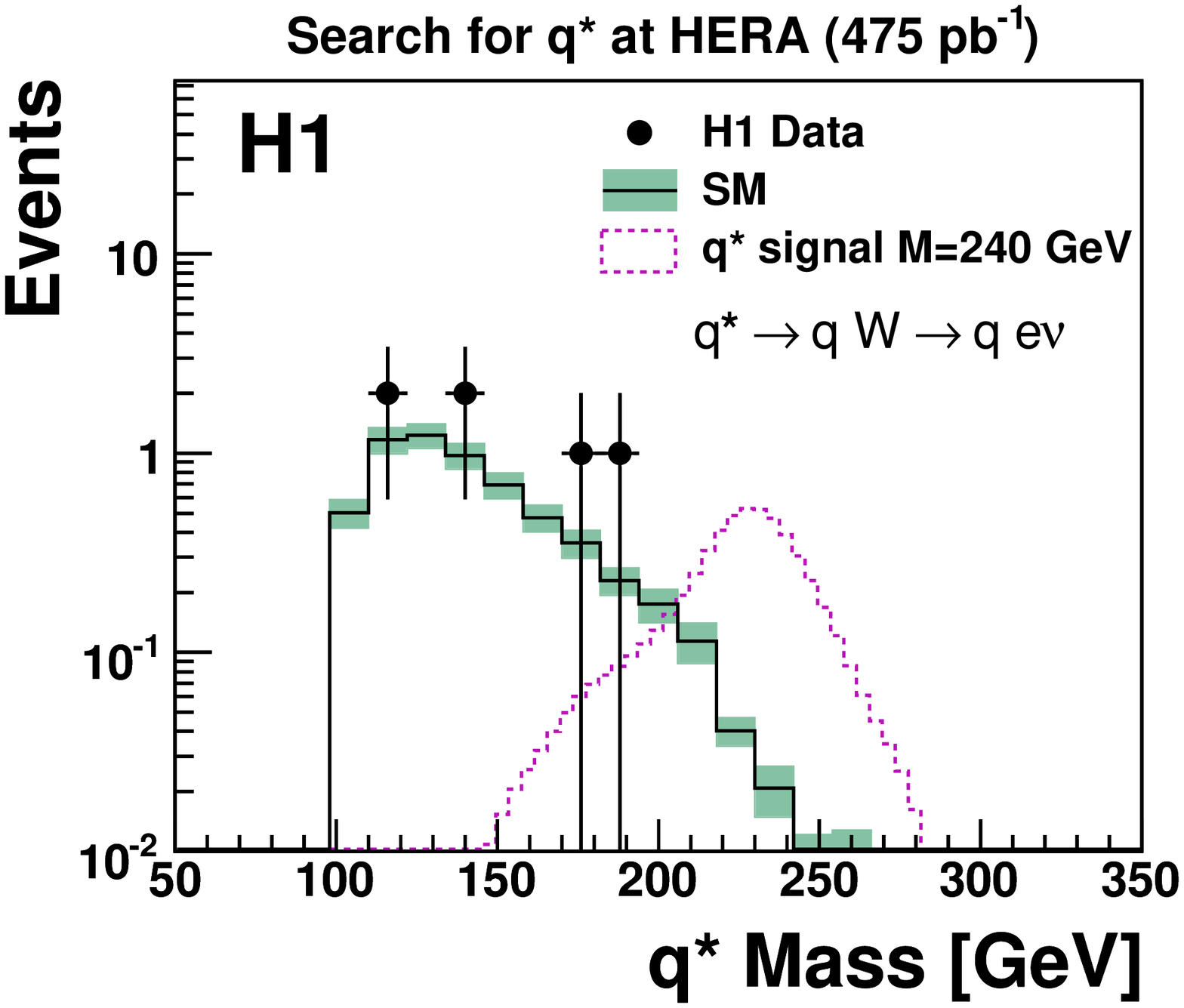}\put(-12,37) {{\bf (c)}} 
\includegraphics[width=.5\textwidth]{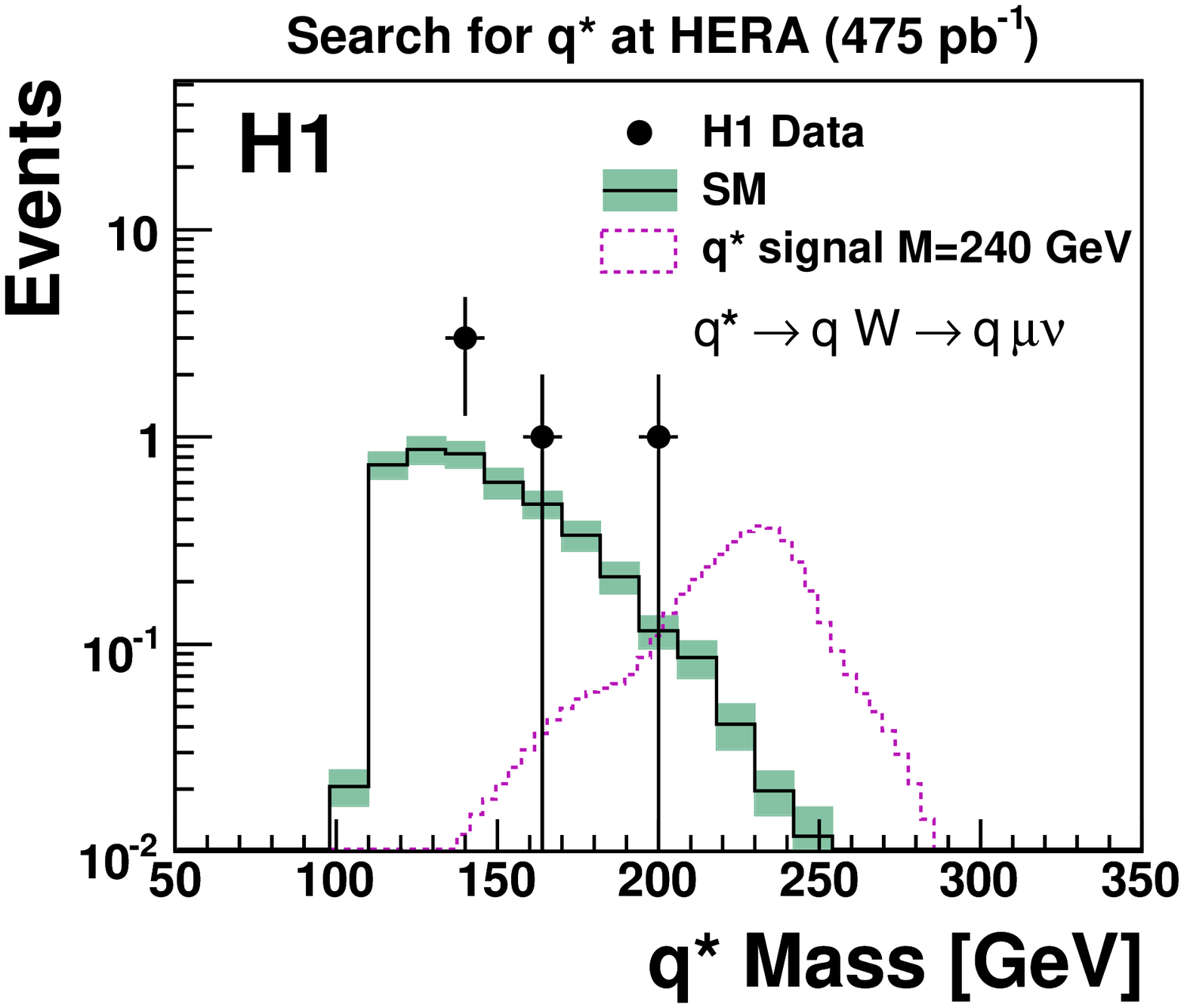}\put(-12,37) {{\bf (d)}}
%\end{center}      
 \caption{Invariant mass distribution of the $q^*$ candidates in the ${q}^{*} {\rightarrow} {q}{\gamma}$ (a), ${q}^{*} {\rightarrow} {q}{W/Z} {\rightarrow} qq\bar{q}$ (b), ${q}^{*} {\rightarrow} q W {\rightarrow} q e \nu$ (c), and  ${q}^{*} {\rightarrow} q W {\rightarrow} q \mu \nu$ (d) search channels. The points correspond to the data and the histograms to the SM expectation after the final selections. The error bands on the SM prediction include model uncertainties and experimental systematic errors added in quadrature.
The dashed line represents the reconstructed mass distribution of MC $q^*$ signal events with $M_{q^*} = 240$~GeV, with an arbitrary normalisation.}
 \label{fig:Mass}  
 \end{figure}

In the search for ${q}^{*} {\rightarrow} {q} W {\rightarrow} qe\nu$, events with $P_T^{\rm{miss}} > 25$~GeV, one electron with $P_T^{e} > 10$~GeV and one jet with $P_T^{\rm{jet}} > 20$~GeV are selected. 
The electron and the jet must be detected in the polar angle range $5^\circ < \theta^{e, \rm{jet}} <  100^\circ$. 
Furthermore, the electron  must be isolated from jets by a minimum distance of  $R > 1$.
The ratio $V_{ap}/V_{p}$ of transverse energy flow anti-parallel and parallel to the hadronic final state~\cite{Adloff:1999ah} is used to suppress photoproduction and NC DIS events. 
Events with $V_{ap}/V_{p} > 0.25$ are rejected.
The invariant mass of the $W$ boson candidate, reconstructed from the four-vectors of the electron and neutrino candidates, is required to be between $55$ and $100$~GeV.
After this selection six data events remain, while $6.0 \pm 0.8$ SM events are expected, of which $82$\% are from single $W$ production. 
The invariant mass of the $q^*$ candidate is calculated from the jet and $W$ candidate four-vectors.
For this calculation, the $W$ candidate four-vector is scaled such that its mass is set to the nominal $W$ boson mass. 
The invariant mass distribution of the $q^*$ candidates and the SM background is presented in figure~\ref{fig:Mass}(c).
The selection efficiency in this channel is  $\sim 20$\% for $M_{q^*} = 120$~GeV, increasing to $30$\% for $M_{q^*} = 260$~GeV.
The total width of the reconstructed $q^*$ mass distribution is $10$~GeV for a generated $q^*$ mass of $120$~GeV, increasing to $20$~GeV for a $q^*$ mass of $260$~GeV.

In the search for ${q}^{*} {\rightarrow} {q} W {\rightarrow} q\mu\nu$, events with $P_T^{\rm{miss}} > 25$~GeV, one muon with $P_T^{\mu} > 10$~GeV and one jet with $P_T^{\rm{jet}} > 15$~GeV are selected. 
The muon and the jet must be detected in the polar angle ranges $5^\circ < \theta^\mu <  100^\circ$ and $5^\circ < \theta^{\rm{jet}} <  160^\circ$, respectively.
To reduce the background contribution from SM $W$ production, the $P_T$ of the jet is required to be larger than $25$~GeV in the polar angle range $\theta^{\rm{jet}} < 60^\circ$. 
A $W$ candidate is reconstructed from the combination of the four-vectors of the muon and neutrino candidates and its mass is required to be larger than $40$~GeV.
After this selection five data events remain, while $4.4 \pm 0.7$ SM events are expected, of which $90$\% are from single $W$ production.
The invariant mass of the $q^*$ candidate is calculated from the jet and $W$ candidate four-vectors.
For this calculation, the $W$ candidate four-vector is scaled such that its mass is set to the nominal $W$ boson mass. 
The invariant mass distribution of the $q^*$ candidates and the SM background is presented in figure~\ref{fig:Mass}(d).
The selection efficiency in this channel is  $\sim 20$\% for $M_{q^*} = 120$~GeV, increasing to $40$\% for $M_{q^*} = 260$~GeV.
The total width of the reconstructed $q^*$ mass distribution is $14$~GeV for a generated $q^*$ mass of $120$~GeV, increasing to $30$~GeV for a $q^*$ mass of $260$~GeV.

\subsection{\boldmath $q e e$ and $q \mu \mu$ Resonance Searches} 

The signature of the ${q}^{*} {\rightarrow} {q} Z {\rightarrow} qee$ and ${q}^{*} {\rightarrow} {q} Z {\rightarrow} q\mu\mu$ decay channels consists of two high $P_T$ leptons and an energetic jet. The production of lepton pairs constitutes the main background contribution from SM processes~\cite{Aaron:2008jh}.

In the search for ${q}^{*} {\rightarrow} {q} Z {\rightarrow} qee$, events with two electrons and one jet of high transverse momenta are selected.
Events are selected with two electrons in the polar angle range $5^\circ < \theta^{e} <  100^\circ$ and transverse momenta larger than $20$ and $10$~GeV. 
A jet  with a transverse momentum larger than $20$~GeV must be detected in the polar angle range $5^\circ < \theta^{\rm{jet}} <  100^\circ$. 
To reduce the background from QED Compton and NC DIS processes, each electron  must be associated to a good quality track also in the forward region ($5^\circ < \theta^{e} < 35^\circ$).
A $Z$ candidate is reconstructed from the combination of the two electrons and its reconstructed mass is required to be compatible with the nominal $Z$ boson mass within $7$~GeV.
After this selection no data event remains, while $0.44 \pm 0.08$ SM events are expected. 

In the search for ${q}^{*} {\rightarrow} {q} Z {\rightarrow} q\mu\mu$, events with two muons and one jet of high transverse momenta are selected.
Events are selected with two muons in the polar angle range $5^\circ < \theta^{\mu} <  160^\circ$ and transverse momenta larger than $15$ and $10$~GeV, respectively. 
A jet with a transverse momentum larger than $20$~GeV must be detected in the polar angle range $5^\circ < \theta^{\rm{jet}} <  100^\circ$. 
A $Z$ candidate is reconstructed from the combination of the two muons and its reconstructed mass is required to be larger than $50$~GeV.
After this selection no data event remains, while $0.87 \pm 0.11$ SM events are expected. 

In both channels, the selection efficiency is $\sim$ $30$\% for events with $m_{q^{*}}>120$ GeV.
The total width of the reconstructed $q^*$ mass distribution in the $q e e$ ($q \mu \mu$) channel is $5$~GeV ($25$~GeV) for a generated $q^*$ mass of $120$~GeV, increasing to $20$~GeV ($30$~GeV) for a $q^*$ mass of $260$~GeV.

\subsection{Systematic Uncertainties}
\label{sec:sys_err}

The following experimental systematic uncertainties are considered:

\begin{itemize}
\item The uncertainty on the electromagnetic energy scale varies between $0.7$\% and $2$\% depending on the polar angle~\cite{trinh}. The polar angle measurement uncertainty is $3$ mrad for electromagnetic clusters.
The identification efficiency of electrons is known with an uncertainty of $3$\%.
\item The efficiency to identify photons is known with a precision of $10$\% for photons with $P_T > 10$~GeV~\cite{trinh}.
\item The scale uncertainty on the transverse momentum of high $P_T$ muons amounts to $2.5$\%. The uncertainty on the reconstruction of the muon polar angle is $3$~mrad.
The identification efficiency of muons is known with an uncertainty of $3$\%.
\item The hadronic energy scale is known within $2$\%~\cite{trinh}. The uncertainty on the jet polar angle determination is $10$ mrad.
\item The uncertainty on the trigger efficiency is $3$\%.
\item The luminosity measurement has an uncertainty of $3$\%.
\end{itemize}

The effect of the above systematic uncertainties on the SM expectation and the signal efficiency is determined by varying the experimental quantities by $\pm 1$ standard deviation in the MC samples and propagating these variations through the whole analysis chain.

Additional model systematic uncertainties are attributed to the SM background MC generators described in section $3$.
An error of $20$\% is attributed to NC DIS, CC DIS and photoproduction processes with at least two high $P_T$ jets. 
It includes uncertainties from the proton distribution functions, from missing higher
order QCD corrections and from hadronisation.
The error on the elastic and quasi-elastic QED Compton cross sections is conservatively estimated to be $5$\%; 
the error on the inelastic QED Compton cross section is $10$\%.
The errors attributed to lepton pair and $W$ production are $3$\% and $15$\%, respectively.
The total error on the SM background prediction is determined by adding the effects of all model and experimental systematic uncertainties in quadrature.

The theoretical uncertainty on the $q^*$ production cross section is dominated by the uncertainty on the scale at which the proton parton densities are evaluated.
It is estimated by varying this scale from $\sqrt{\hat{s}}/2$ to $2\sqrt{\hat{s}}$.
The resulting uncertainty depends on the $q^*$ mass and is $5$\% at $M_{q^*} = 100$~GeV, increasing to $12$\% at $M_{q^*} = 300$~GeV.

%%%%%%%%%%%%%%%%%%%%%%%%%%%%%%%%%%%%%%%%%%%%%%%%%%%%%%%%%%%%%%%%%%%%%%%%%
\section{Interpretation and Limits}
%%%%%%%%%%%%%%%%%%%%%%%%%%%%%%%%%%%%%%%%%%%%%%%%%%%%%%%%%%%%%%%%%%%%%%%%%

The event yields observed in all decay  channels are in agreement with the corresponding SM expectations and are summarised in table~\ref{tab:qstaryields}. 
The SM predictions are dominated by photoproduction and NC DIS for searches in the $q\gamma$ and $q{q}{\bar{q}}$ channels and by SM $W$ production in the $qe\nu$ and $q\mu\nu$ channels. 
The observed invariant mass distributions are in agreement with those of the SM background as shown in figure~\ref{fig:Mass}. 
No data events are observed in channels corresponding to leptonic decays of the $Z$ boson, in agreement with the low SM expectations.

Since no evidence for the production of excited quarks is observed, upper limits on the $q^*$ production cross section and on the model parameters are derived as a function of the mass of the excited quark. 
Limits are presented at the $95$\% confidence level (CL) and are obtained from the mass spectra using a modified frequentist approach which takes statistical and systematic uncertainties into account~\cite{Junk:1999kv}.

\begin{figure}[htbp] 
  \begin{center}
\includegraphics[width=.5\textwidth]{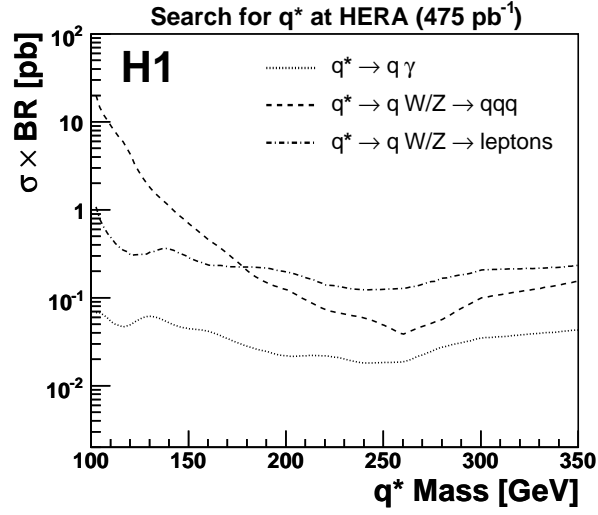}
   \end{center}
  \caption{Upper limits at $95$\% CL on the product of the $q^*$ cross section and decay branching ratio, $\sigma \times$~BR, in the three types of final states for $q^*$ events as a function of the excited quark mass. The $q^*$ decay channels into the $W$ and $Z$ bosons are combined. Values above the curves are excluded. }
\label{fig:LimitSBR}  
\end{figure}

\begin{figure}[htbp]
  \begin{center}
\includegraphics[width=.5\textwidth]{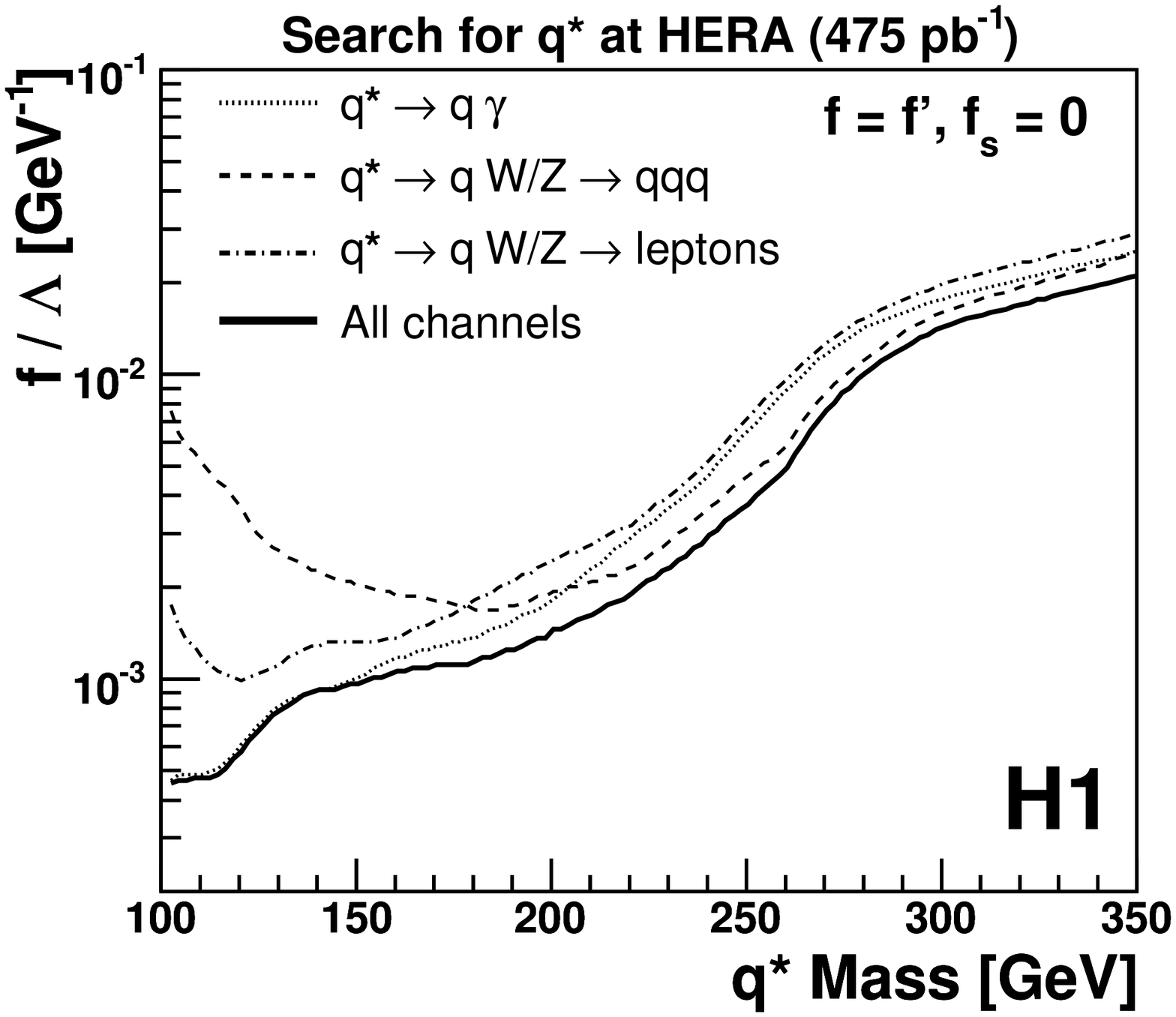}\put(-10,35){{\bf (a)}}
\includegraphics[width=.5\textwidth]{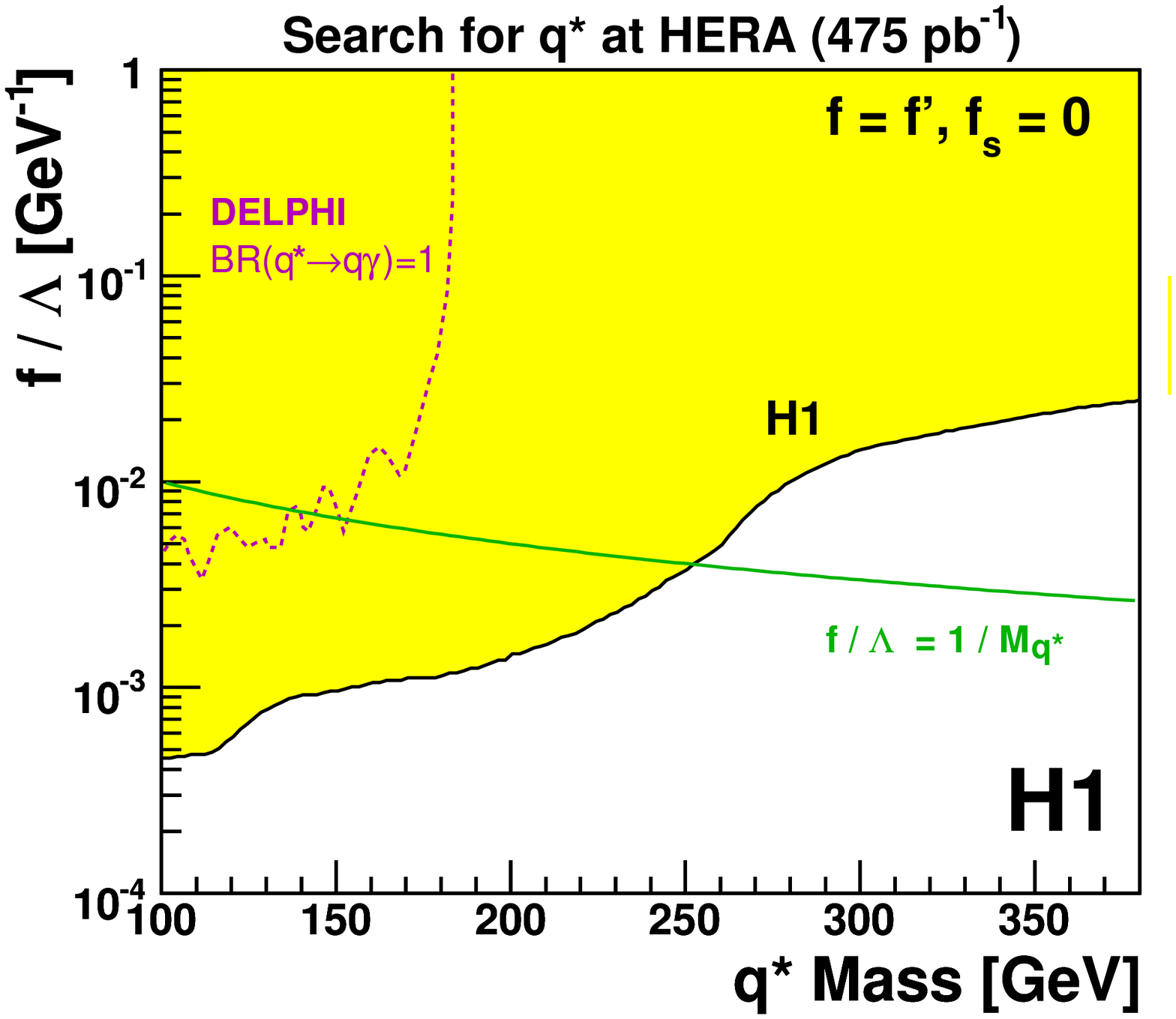}\put(-10,35){{\bf (b)}}
  \end{center}
  \caption{Exclusion limits  at $95$\% CL on $f/\Lambda$  as a function of the mass of the excited quark with the assumptions $f = f'$ and $f_s =0$.
The individual contributions of the $q^*$ decay channels are presented in (a).  
Values of the couplings above the curves are excluded.
The excluded domain based on  all H1 $e^\pm p$ data is represented in (b) by the shaded area.
It is compared to the exclusion limit obtained at LEP by the DELPHI Collaboration~\protect{\cite{Abreu:1998jw}} (dashed line), assuming that the branching ratio of the $q^* \rightarrow q \gamma$ is equal to $1$.
The curve $f/\Lambda = 1/M_{q^*}$ is indicated in (b).}
\label{fig:LimitCoupling}  
\end{figure}

Upper limits on the product of the $q^*$ production cross section and of the $q^*$ final state branching ratio are shown in figure~\ref{fig:LimitSBR}. 
The analysed $q^*$ decays into $W$ and $Z$ gauge bosons are combined.
The resulting limit on $f/\Lambda$ after combination of all decay channels is displayed as a function of the $q^*$ mass in figure~\ref{fig:LimitCoupling}, for $f_s = 0$ and the conventional assumption $f = f'$. 
Under the assumption $f/\Lambda = 1/M_{q^*}$ excited quarks with masses below $252$~GeV are excluded. 
The individual limits from different $q^*$ decay channels are also shown in figure~\ref{fig:LimitCoupling}(a).
At low mass, the combined limit on $f/\Lambda$ is dominated by the $q^* {\rightarrow} q \gamma$ channel, while the \mbox{$q^* \rightarrow q W$} and $q^* \rightarrow q Z$ channels start to contribute for masses above $150$~GeV and dominate for masses above $200$~GeV.
These new results extend the previously published limits by H1~\cite{Adloff:2000gv} and ZEUS~\cite{Chekanov:2001xk} by a factor of two to five in $f/\Lambda$. Constraints on $q^*$ masses beyond the HERA kinematic limit are set.
Excited quarks with masses below $380$~GeV are excluded for coupling values $f/\Lambda > 0.03$~GeV$^{-1}$ (see figure~\ref{fig:LimitCoupling}(b)).
The exclusion limit obtained at LEP by the DELPHI Collaboration~\protect{\cite{Abreu:1998jw}}  assuming that the branching ratio of the $q^* \rightarrow q \gamma$ is equal to $1$ is also shown in figure~\ref{fig:LimitCoupling}(b) and is considerably weaker than the present result.

\begin{figure}[htbp] 
  \begin{center}
\includegraphics[width=.5\textwidth]{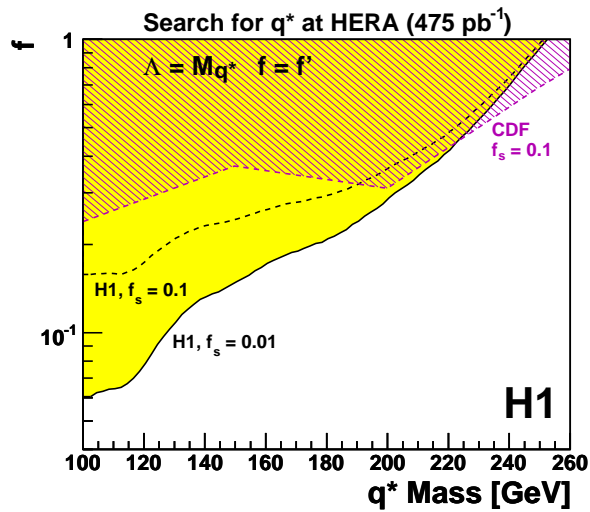}
   \end{center}
  \caption{Exclusion limits at $95$\% CL on the coupling $f$ as a function of the mass of the excited quark, assuming $\Lambda = M_{q^*}$ and $f = f'$ for two different values of $f_s$.
Values above the curves are excluded. 
Also shown is the exclusion limit obtained at the Tevatron by the CDF Collaboration~\protect{\cite{Abe:1993sv}} derived for $f_s =0.1$ (hatched area).}
\label{fig:Limit_fs}  
\end{figure}

The sensitivity of this analysis to non-zero values of $f_s$ and its complementarity to results obtained at the Tevatron is also studied.
The limit on $f$ obtained for two example values of $f_s$ and under the assumptions $f = f'$ and $\Lambda = M_{q^*}$ is presented in figure~\ref{fig:Limit_fs}. 
This limit is derived using the $\gamma$, $W$ and $Z$ decay channels of excited quarks. 
The $q^* \rightarrow qg$ decay channel gives rise to a dijet resonance.
It was verified in a complementary analysis~\cite{GS_H2} that no resonance from two high $P_T$ jets is observed in the present data.
Due to the overwhelming dijet SM background, the total limit on $q^*$ production is not improved if the $q^* \rightarrow qg$ decay channel is included.
The present limit is compared to the limit obtained by the CDF Collaboration for $f_s = 0.1$ and the same assumptions~\cite{Abe:1993sv}.
For $f_s < 0.1$ and for $M_{q^*} < 190$~GeV, the present analysis probes a domain not excluded by Tevatron experiments.
In the case $f = f' = f_s = 1$ and $\Lambda = M_{q^*}$, Tevatron experiments are able to exclude excited quark masses up to $870$~GeV~\cite{Abazov:2003tj,Aaltonen:2008dn}.

%%%%%%%%%%%%%%%%%%%%%%%%%%%%%%%%%%%%%%%%%%%%%%%%%%%%%%%%%%%%%
\section{Conclusion}
%%%%%%%%%%%%%%%%%%%%%%%%%%%%%%%%%%%%%%%%%%%%%%%%%%%%%%%%%%%%%

A search for the production of excited quarks is performed using the full $e^{\pm}p$ data sample collected by the H1 experiment at HERA with an integrated luminosity of $475$~pb$^{-1}$. 
The excited quark decay channels ${q}^{*} {\rightarrow} {q}{\gamma}$,  ${q}^{*} {\rightarrow} {q}{Z}$ and ${q}^{*} {\rightarrow} {q}{W}$ with subsequent hadronic or leptonic decays of the $W$ and $Z$ bosons are considered and no indication of a signal is found.
Improved limits on the production cross section of excited quarks are obtained. 
Within gauge mediated models, an upper limit on the coupling $f/\Lambda$ as a function of the excited quark mass is established for the specific relations $f = f'$ and $f_s =0$.
For $f/\Lambda=1/M_{q^*}$ excited quarks with a mass below $252$~GeV are excluded at $95$\% confidence level.
The results presented in this paper extend previously excluded domains at HERA and LEP and are complementary to $q^*$ searches performed at the Tevatron.

%%%%%%%%%%%%%%%%%%%%%%%%%%%%%%%%%%%%%%%%%%%%%%%%%%%%%%%%%%%%
\section*{Acknowledgements}

We are grateful to the HERA machine group whose outstanding
efforts have made this experiment possible. 
We thank the engineers and technicians for their work in constructing 
and maintaining the H1 detector, our funding agencies for financial 
support, the DESY technical staff for continual assistance and the 
DESY directorate for the hospitality which they extend to the non-DESY 
members of the collaboration.

%%%%%%%%%%%%%%%%%%%%%%%%%%%%%%%%%%%%%%%%%%%%%%%%%%%%%%%%%%%%
\begin{flushleft}

\end{flushleft}

%%%%%%%%%%%%%%%%%%%%%%%%%%%%%%%%%%%%%%%%%%%%%%%%%%%%%%%%%%%%%%%%%%%%%%%%%%%%%%
\clearpage

\begin{table}[]
\begin{center}
\begin{tabular}{l c c c}
\multicolumn{4}{c}{{\bf H1 Search for \begin{boldmath}$q^*$\end{boldmath} at HERA (\begin{boldmath}$475$\end{boldmath} pb\begin{boldmath}$^{-1}$\end{boldmath})}}\\
\hline
Channel & ~Data~ & SM & Signal Efficiency [\%] \rule[-6pt]{0pt}{19pt}\\
\hline
${q}^{*} {\rightarrow}q \gamma$  & $44$ & $46 \pm \;\;\;4\;\; \pm \;7$~ & $35$ -- $45$\\
${q}^{*} {\rightarrow} q W/Z {\rightarrow} qq\bar{q}~~~$ & $341$ & $326 \pm \;48\;\; \pm 62$~~ &  ~~$5$ -- $55$\\

${q}^{*} {\rightarrow} q W {\rightarrow} q e \nu$ & $6$ & $6.0 \pm 0.2\;\; \pm 0.8$ & $20$ -- $30$\\
${q}^{*} {\rightarrow} q W {\rightarrow} q \mu \nu$ & $5$ & $4.4 \pm  0.2\;\; \pm 0.7$ & $20$ -- $40$\\

 ${q}^{*} {\rightarrow} q Z {\rightarrow} qee$ & $0$ & $0.44 \pm 0.06 \pm 0.04$ & $15$ -- $30$ \\
 ${q}^{*} {\rightarrow} q Z {\rightarrow} q\mu\mu$ & $0$ & $0.87 \pm 0.10 \pm 0.04$ & $15$ -- $30$\\
\hline
\end{tabular}
\end{center}

\caption{
Observed and predicted event yields for the studied $q^*$ decay channels.
%  The analysed data sample corresponds to an integrated luminosity of $475$~pb$^{-1}$.
  The first and second errors on the SM predictions correspond to experimental and model systematic errors, respectively.
  Typical selection efficiencies obtained from MC studies for $q^*$  masses ranging from $120$ to $260$~GeV are also indicated.
}
\label{tab:qstaryields}
\end{table}

%%%%%%%%%%%%%%%%%%%%%%%%%%%%%%%%%%%%%%%%%%%%%%%%%%%%%%%%%%%%%%%%%%%%%%%%%%%%%%
\clearpage

\end{document}